\documentclass{mem}
\usepackage{natbib}
\usepackage{txfonts}
\usepackage{balance}
\usepackage{graphicx}
\usepackage[a4paper]{hyperref}
\idline{75}{282}
\begin{document}
\def\teff{$T\rm_{eff }$}
\def\kms{$\mathrm {km s}^{-1}$}

\title{Low-frequency radio observations of the galaxy cluster CIZA~J2242.8+5301
}

   \subtitle{}

   \author{R.~J. van Weeren\inst{1}
         \and H.~T.~Intema\inst{4}
         \and H.~J.~A. R\"ottgering\inst{1}
         \and M.~Br\"uggen\inst{2}
         \and M.~Hoeft\inst{3}
          }

  \offprints{R.~J. van Weeren}

   \institute{Leiden Observatory, Leiden University,
              P.O. Box 9513, NL-2300 RA Leiden, The Netherlands\\
              \email{rvweeren@strw.leidenuniv.nl}
                 \and Jacobs University Bremen, P.O. Box 750561, 28725 Bremen, Germany
                 \and Th\"uringer Landessternwarte Tautenburg, Sternwarte 5, 07778, Tautenburg, Germany
                 \and National Radio Astronomy Observatory, 520 Edgemont Road, Charlottesville, VA 22903-2475, USA
                 }

\authorrunning{van Weeren}

\titlerunning{Radio observations of CIZA~J2242.8+5301}

\abstract{  
Some disturbed galaxy clusters host diffuse elongated radio sources, also called radio relics. It is proposed that these relics trace shock waves in the intracluster medium (ICM). Within the shock waves, generated by cluster merger events, particles are accelerated to relativistic energies, and in the presence of a magnetic field synchrotron radiation will be emitted. 
CIZA~J2242.8+5301 is a disturbed galaxy cluster hosting complex diffuse radio emission, including a so-called double radio relic. Here we present new Giant Metrewave Radio Telescope (GMRT) radio observations of CIZA~J2242.8+5301 at 325 and 150~MHz. We detect the double radio relic at 150 and 325 MHz. The very deep 150~MHz image reveals the presence of large-scale diffuse emission between the two radio relics.    
}
\maketitle{}

\section{Introduction}
Galaxy clusters grow by mergers with other clusters or galaxy groups, as well as through the accretion of gas from the intergalactic medium (IGM). Both these two processes shock the ICM. It has been proposed that within these shocks particles can be accelerated to highly relativistic energies by the diffusive shock acceleration (DSA) mechanism \citep{1977DoSSR.234R1306K, 1977ICRC...11..132A, 1978MNRAS.182..147B, 1978MNRAS.182..443B, 1978ApJ...221L..29B, 1983RPPh...46..973D, 1987PhR...154....1B, 1991SSRv...58..259J, 2001RPPh...64..429M}. It is thought that radio relics, elongated steep-spectrum radio sources \citep[e.g.,][]{1991A&A...252..528G, 2006Sci...314..791B,2008A&A...486..347G,2011ApJ...727L..25B}, trace these shock waves generated by cluster merger events \citep{1998A&A...332..395E, 2001ApJ...562..233M}. 
\begin{figure*}
\resizebox{\hsize}{!}{\includegraphics[angle =90, trim =0cm 0cm 0cm 0cm,width=1.0\textwidth, clip=true]{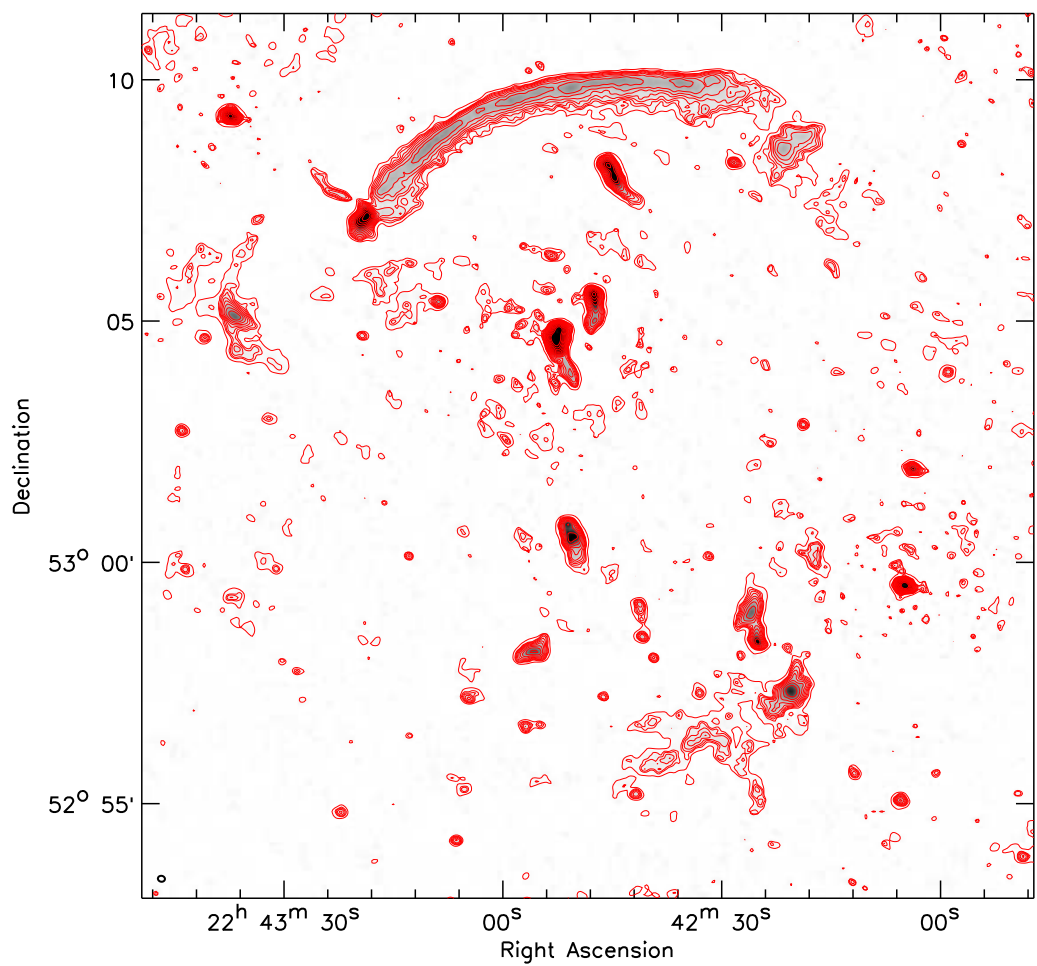}}
\caption{\footnotesize GMRT 325~MHz image made with robust weighting set to $-1.0$ \citep{briggs_phd}. The image has a rms noise of 59~$\mu$Jy~beam$^{-1}$. Contour levels are drawn at $\sqrt{[1, 2, 4, 8, \ldots]} \times 4\sigma_{\mathrm{rms}}$. The beam size is $8.7\arcsec \times 7.4\arcsec$ and shown in the bottom left corner of the image. A thinner contour from an image with robust weighting set to $0.5$, with a resolution of $11.5\arcsec \times 9.8\arcsec$, is drawn at a level of $0.2$~mJy~beam$^{-1}$. 
}
\label{fig:325}
\end{figure*}

CIZA~J2242.8+5301 is a disturbed galaxy cluster located at $z=0.1921$ \citep{2007ApJ...662..224K}. The cluster hosts a large double radio relic system, as well as additional large-scale diffuse radio emission \citep{2010Sci...330..347V}. The spectral index ($\alpha$) across the bright northern relic steepens systematically in the direction of the cluster center, across the full length of the narrow relic. This is expected for outwards moving shock waves, with synchrotron and inverse Compton losses behind the shock front. Here we present new low-frequency GMRT radio images at 150 and 325~MHz, which complement our previous higher frequency observations at 610 and 1400~MHz \citep{2010Sci...330..347V}. 

\section{Observations}
We carried out radio observations with the GMRT of  CIZA~J2242.8+5301 at 150 and 325 MHz. The total bandwidth was 6 and 32~MHz at 150 and 325~MHz, respectively. The 325~MHz data were reduced in a standard way using the NRAO Astronomical Image Processing System (AIPS) package. A few bright sources were removed using the ``peeling'' scheme \citep[e.g.,][]{2004SPIE.5489..817N}. The data were affected by strong radio frequency interference (RFI), preventing the use of the shortest baselines.

For the 150~MHz data the ionospheric calibration scheme from \cite{2009A&A...501.1185I} was employed. In addition, RFI was fitted and subtracted from the data using the technique described by \cite{2009ApJ...696..885A} which was implemented in  Obit \citep{2008PASP..120..439C}. 

For both datasets we used the 
polyhedron method \citep{1989ASPC....6..259P, 1992A&A...261..353C} 
for making the images to minimize the effects 
of non-coplanar baselines.

\begin{figure}
\resizebox{\hsize}{!}{\includegraphics[angle =90, trim =0cm 0cm 0cm 0cm,width=0.5\textwidth, clip=true]{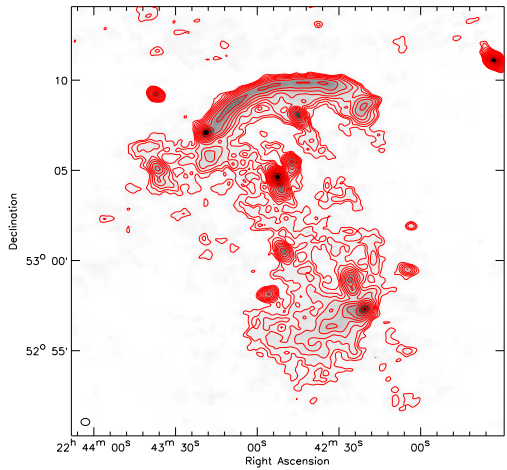}}
\caption{\footnotesize GMRT 150~MHz image. The image has a rms noise of 1.0~mJy~beam$^{-1}$, and the resolution is $30\arcsec~\times~24\arcsec$. Contour levels are drawn at $\sqrt{[1, 2, 4, 8, \ldots]} \times 4\sigma_{\mathrm{rms}}$. 
}
\label{fig:150}
\end{figure}

\section{Results}
The 325~MHz image displays both the northern and southern relic. At 325 MHz we also detect a patch of diffuse emission located just south of the bright tailed-radio source at the eastern end of the northern giant relic. In addition, the diffuse source to the southeast of the tailed radio source is detected. These sources could be radio relics tracing additional shock structures. The 325 MHz image reveals none of the large-scale diffuse emission between the southern and northern giant relics, which was previously found in a deep Westerbork Synthesis Radio Telescope (WSRT) 1.4~GHz image \citep{2010Sci...330..347V}.  We attribute this to the fact that most of the short baseline had to be ``flagged'' because of RFI.  At 150~MHz we detect the large-scale diffuse component between the northern and southern relics, confirming its existence. Interestingly, the width of the northern relic is much wider than in our 1.4~GHz WSRT map, consistent with a very steep
spectral index ($\alpha \lesssim -2$) in the region at the back of the shock due to synchrotron
losses having severely reduced the emission at higher frequencies.

\section{Conclusions}
We detect both the northern and southern radio relics in the galaxy cluster \object{CIZA~J2242.8+5301} with the GMRT at 150 and 325~MHz . The 150~MHz image reveals the presence of additional diffuse emission throughout the cluster with a total extent of about 3~Mpc, confirming the results from our a previous WSRT 1.4~GHz observation. In a future paper we will present a more detailed spectral analysis of the radio emission in the cluster.

\begin{acknowledgements}
We thank the staff of the GMRT who have made these 
observations possible. The GMRT is run by the National 
Centre for Radio Astrophysics of the Tata Institute of 
Fundamental Research. The Westerbork Synthesis Radio 
Telescope is operated by ASTRON (Netherlands 
Institute for Radio Astronomy) with support from the 
Netherlands Foundation for Scientific Research (NWO). 

RJvW acknowledges funding from the Royal  
Netherlands Academy of Arts and Sciences. MB acknowledges support by the research group FOR 1254 funded by the Deutsche Forschungsgemeinschaft.
\end{acknowledgements}

\bibliographystyle{aa}

\bibliography{../filaments/ref_filaments}

\end{document}